\documentclass{aa}
\usepackage{graphicx}
\usepackage{txfonts}
\def\XMM{{\sl XMM-Newton}}
\def\Rosat{{\sl Rosat}}

\begin{document}
   \title{\XMM{} spectra of hard spectrum
\Rosat{} AGN: X--ray absorption and optical reddening}

   \subtitle{}

   \author{F.J. Carrera \inst{1}
          \and
           M.J. Page \inst{2}
          \and
	  J.P.D. Mittaz \inst{3}
          }


   \institute{Instituto de F\'\i{}sica de Cantabria (CSIC-UC), Avenida de los
Castros, 39005 Santander (Spain)\\
              \email{carreraf@ifca.unican.es}
         \and
	     Mullard Space Science Laboratory-University College London, Surrey
RH5 6NT, United Kingdom\\
             \email{mjp@mssl.ucl.ac.uk}
         \and
	     University of Huntsville, Alabama, United States
	      \email{mittazj@email.uah.edu}
             }

   \date{Received November 15, 2003; accepted March 3, 2004}

   \abstract{ We present the \XMM{} spectra of three low-redshift intermediate
   Seyferts (one Sy1.5, and two Sy1.8), from our survey of hard spectrum
   \Rosat{} sources. The three AGN are well fitted by absorbed powerlaws, with
   intrinsic nuclear photoelectric absorption from column densities between 1.3
   and 4.0 $\times 10^{21}$~cm$^{-2}$.  In the brightest object the X--ray
   spectrum is good enough to show that the absorber is not significantly
   ionized.  For all three objects the powerlaw slopes appear to be somewhat
   flatter ($\Gamma\sim1.3-1.6$) than those found in typical unabsorbed
   Seyferts.  The constraints from optical and X--ray emission lines imply that
   all three objects are Compton-thin. For the two fainter objects, the
   reddening deduced from the optical broad emission lines in one of them, and
   the optical continuum in the other, are similar to those expected from the
   X--ray absorption, if we assume a Galactic gas-to-dust ratio and reddening
   curve.  The broad line region Balmer decrement of our brightest object is
   larger than expected from its X--ray absorption, which can be explained
   either by an intrinsic Balmer decrement with standard gas-to-dust ratio, or
   by a $>$Galactic gas-to-dust ratio.
   These $\ge$~Galactic ratios of extinction to photoelectric
   absorption cannot extend to the high redshift, high luminosity, broad line
   AGN in our sample, because they have column densities $>10^{22}$~cm$^{-2}$,
   and so their broad line regions would be totally obscured. This means that
   some effect (e.g., luminosity dependence, or evolution) needs to be present
   in order to explain the whole population of absorbed AGN.
   }

   \titlerunning{X--ray absorption and optical reddening in hard \Rosat{}
   AGN}
   \maketitle
%

\section{Introduction}

According to the unified model for Active Galactic Nuclei (AGN)(Antonucci
1993), broad-line Seyferts (type 1) and narrow-line (type 2) Seyferts are
intrinsically the same type of object but are viewed with different
orientations to our line of sight. In this model the central engine of the AGN,
and the high velocity clouds that produce the broad optical and UV emission
lines, are surrounded by a thick torus of dust and cool, molecular gas. In type
1 objects we have a direct view of the central engine and the broadline clouds,
whereas in type 2 objects the torus blocks our line of sight to these
regions. Such a torus has a large photoelectric opacity at soft X--ray
energies, which explains why type 1 Seyferts have steep X--ray spectra with
little absorption (Nandra \& Pounds 1994), and type 2 Seyferts have absorbed
X--ray spectra (Smith \& Done 1996). Narrow emission lines are produced in
more
distant gas clouds on scales larger than the dusty torus, and so can be seen in
both types of object.  Objects in which the broad lines are attenuated, but not
completely obscured, are called intermediate Seyferts, and are assigned
classifications ranging from 1.5-1.9.

However, even in the type 1 objects, an extra absorption component from ionized
gas is often seen in the X--ray band (George et al.  1998).  This ionized gas,
the ``warm absorber'', appears to be distributed throughout the BLR and NLR of
Seyfert galaxies, and in type 2 objects it can scatter nuclear radiation into
our line of sight. The warm absorber has a much lower opacity to soft X--rays
than cold gas, but if it contains a significant element of dust (e.g. Brandt,
Fabian \& Pounds 1996) it could produce considerable extinction at optical
and
ultraviolet wavelengths.

An understanding of absorption in AGN is extremely important.  For example,
unified AGN models for the X--ray background (XRB) (see e.g., Setti
\& Woltjer 1989, for an early proposal, and Gilli, Salvati \& Hasinger
2001, for a late development) explain the spectrum of the XRB by the
superposition of spectra of AGN with various degrees of
absorption. Under this scheme, soft spectrum X--ray sources should be mostly
type 1 AGN, while hard spectrum X--ray sources would be predominantly type 2
AGN. In the 1990s this matched the observations quite well, because \Rosat{}
surveys in the soft band were dominated by type 1 AGN, (e.g. Mason et al. 2000,
Lehman et al. 2001) while surveys selected in harder bands with {\sl BeppoSAX}
(Fiore et al. 1999) were much richer in type 2 AGN.

Despite these early successes, recent developments have put in jeopardy the
identity between optical type 1 and X--ray unabsorbed objects, on one hand, and
optical type 2 and X--ray absorbed objects, on the other. For example,
identifications of our survey of \Rosat{} sources with hard spectra (Page,
Mittaz \& Carrera 2000, 2001) produced mostly type 1 AGN, contrary to the
expectations of the unified model. Other examples of X--ray absorbed type 1
objects have been found by 
%
Akiyama et al. (2000) using {\sl ASCA} data, and Mainieri et al. (2002) and
Page et al. (2003) using \XMM{} data. In principle, high gas-to-dust ratios in
the X--ray absorbing gas (perhaps due to dust sublimation close to the central
X--ray source, Granato, Danese \& Franceschini 1997), or large dust grains
(Maiolino, Marconi \& Oliva 2001), could give rise to high levels of X--ray
absorption, without much optical obscuration.

In the opposite sense, Pappa et al. (2001) have found several examples of type
2 AGN with very little or no X--ray absorption. Panessa \& Bassani (2002)
estimate that 10-30\% of Seyfert 2 galaxies have this property.  One striking
example is H1320+551 (Barcons, Carrera \& Ceballos 2003), a Seyfert 1.8/1.9
galaxy with strong optical (BLR and NLR) obscuration, but without any
corresponding X--ray absorption from cold gas. The high quality of their \XMM{}
data allows these authors to rule out a warm absorber in this source, which
leads them to the conclusion that the BLR is intrinsically reddened in this
object: its Sy 1.8/1.9 appearance cannot arise from obscuration of a Seyfert 1
spectrum.

The situation is therefore complex.  Possible explanations include
Compton thick obscuration which could suppress completely the nuclear emission
below 10~keV. This spectral range could then be filled by X--rays scattered off
the warm absorber, or by extranuclear emission, which would not have in
principle a particularly hard or absorbed spectrum. This could result in
optically obscured type 2 AGN which appear to be absorption free at X--ray
energies. Such a model can in principle be tested, since Bassani et al. (1999)
have developed a diagnostic diagram that permits identifying Compton thick
sources, as those with high equivalent width Fe emission lines (originating in
fluorescence in the torus material), and low 2-10~keV to [OIII] flux ratio.
This is based on the observation that [OIII] originates in the NLR, outside the
torus, and thus in principle [OIII] should be free of obscuration. Neither
H1320+551, nor any of the sources discussed in Pappa et al. (2001), lie in the
Compton thick region of this diagram. They represent therefore genuine
mismatches between optical and X--ray classifications, at odds with the unified
AGN model.

Here, we analyze optical and \XMM{} spectroscopic data on three AGN
(RXJ133152.51\-+111643.5, RXJ163054.25\-+781105.1, and RXJ213807.61\--423614.3)
from the sample of Page, Mittaz \& Carrera (2001). Of the objects in this
sample which show broad optical emission lines, these three had the highest
X--ray fluxes.  All three show strong signs of absorption in their \Rosat{}
spectra. In section \ref{OSp} we present their optical spectra, finding
evidence for optical obscuration in at least two of them. We then analyze their
\XMM{} spectra in section \ref{XSp}, and in particular we measure their
intrinsic X--ray absorption. The differences between the levels of optical
obscuration and X--ray absorption are discussed in section \ref{discussion}, as
well as a comparison of the X--ray spectral properties of our sources with
respect to those of other samples at similar flux levels. Finally, in section
\ref{discussion} we summarize our results.

For brevity, we will refer to the three sources using truncated versions of
their names (i.e. RXJ1331, RXJ1630 and RXJ2138) in the text.  We have used the
currently fashionable values of $H_0=70$ km s$^{-1}$ Mpc$^{-1}$,
$\Omega_{m}=0.3$, and $\Omega_{\Lambda}=0.7$, throughout this paper.

\section{Optical spectra}
\label{OSp}

The three sources in this study were taken from the Page, Mittaz \& Carrera
(2001) sample of sources with hard \Rosat{} spectra. One of them (RXJ2138) was
identified as an AGN at $z=0.019$ during the follow-up programme for that
project. For this optical spectrum we make use of the photon statistical errors
which were propagated through the data reduction. The other two objects
(RXJ1331 and RXJ1630) were identified during the RIXOS project (Mason et
al. 2000), as AGN at $z=0.090$ and $z=0.358$ respectively, and we have taken
the optical spectra from the archive of that project. For these two spectra,
statistical errors are not available, so we have estimated the statistical
uncertainties from the dispersion of the data around a straight line in an
emission-line-free continuum region. We have estimated the confidence intervals
on the fitted parameters using the standard $\Delta\chi^2$ technique. Given the
low resolution of the spectra, in all the fits the relative central wavelengths
of the lines with respect to $H\beta$ have been fixed to their rest values. We
have assumed that for each object, all the narrow lines have the same width,
and similarly for the broad lines, to keep the number of independent parameters
to a minimum.

The optical reddening can be calculated using $E(B-V)=2.07\log(({\rm
H\alpha/H\beta})/({\rm H\alpha/H\beta})_{\rm intrinsic})$ (Osterbrock 1989),
where the intrinsic Balmer decrements are 3 for the NLR and 3.43$\pm$0.19 for
the BLR (see appendix \ref{appendix}). Assuming a standard gas-to-dust ratio,
the total Hydrogen (HI+H$_2$) column density is then given by $N_H=5.8\times
10^{21}E(B-V)$~cm$^{-2}$ (Bohlin, Savage \& Drake 1978). We have added in
quadrature an additional 10\% uncertainty on the Balmer decrements, to take
into account possible relative flux calibration differences over large ranges
of the spectra.

We have also compared the broad-band optical spectral shape of RXJ1630 (for
which no BLR Balmer decrement can be obtained) with that of the composite QSO
spectrum from Francis (2003), with some contribution from a galaxy continuum
(in our case we have tried an ES0 from the Coleman, Wu \& Weedman 1980
model). See appendix \ref{appendix} for a discussion on this technique.

A more detailed analysis of the optical spectra of the three sources
follows. The best fit parameters are summarized in Table \ref{OSpResults}. 

   \begin{table*}
      \caption[]{Summary of parameters from fits to optical spectra. All errors
quoted are 1$\sigma$ for 1 d.o.f. The [OIII] line referred to is at a rest
wavelength of 5007\AA. The columns marked  $E(B-V)$ give the
optical reddening from the Balmer decrement, except for the one marked *
which is from broad-band properties(see text), with the Galactic reddening
subtracted in all cases.}
         \label{OSpResults}
     
         \begin{tabular}{llrrrrrrrrrr}
            \hline
            \noalign{\smallskip}
       &    & \multicolumn{5}{c}{Narrow lines} && 
               \multicolumn{4}{c}{Broad lines} \\
\cline{3-7}\cline{9-12}
Source & $z$& $\sigma$ & \multicolumn{3}{c}{Intensity} &
$E(B-V)$ &&
              $\sigma$ & \multicolumn{2}{c}{Intensity} & $E(B-V)$ \\
       &    & (\AA) & \multicolumn{3}{c}{($10^{-15}$ erg cm$^{-2}$
s$^{-1}$)}& (mag) &&
              (\AA) & \multicolumn{2}{c}{($10^{-15}$ erg cm$^{-2}$
s$^{-1}$)}&(mag)\\
\cline{4-6}\cline{10-11}
       &    &      & H$\beta$ & [OIII] & H$\alpha$ &
&&
                   & H$\beta$ & H$\alpha$ \\

            \noalign{\smallskip}
            \hline
            \noalign{\smallskip}

RXJ1331 & $0.090\pm0.004$ & 
    6.5  & 9.1$\pm$0.3   & 101.5$^{+0.2}_{-0.4}$ &
31.6$^{+0.2}_{-0.4}$  & $0.10\pm0.04$ &&
   75.2  & 20$^{+1.2}_{-1.1}$                    &
239.74$^{+0.03}_{-0.05}$      & $1.10\pm0.07$\\  
RXJ2138 & $0.019\pm0.004$   &
    6.3  & 1.14$^{+0.15}_{-0.17}$ & 2.1$^{+0.15}_{-0.16}$&
3.4$^{+0.11}_{-0.17}$ & $-0.05\pm0.12$ &&
   83    & 3.9$^{+0.5}_{-0.6}$ & 25.8$^{+0.3}_{-0.6}$                   
& $0.60\pm0.14$\\
RXJ1630 & $0.358\pm0.003$  &
    7.9  & 0.11$\pm$0.03 & 1.36$^{+0.03}_{-0.02}$& -                
    & - &&
  107    & 1.98$^{+0.10}_{-0.09}$              & -                      
& $0.00\pm$0.17*\\

            \noalign{\smallskip}
            \hline
         \end{tabular}

\end{table*}

\subsection{RXJ133152.51+111643.5}
\label{278-010OSp}

\begin{figure}
   \centering
   \includegraphics[angle=-90,width=8cm]{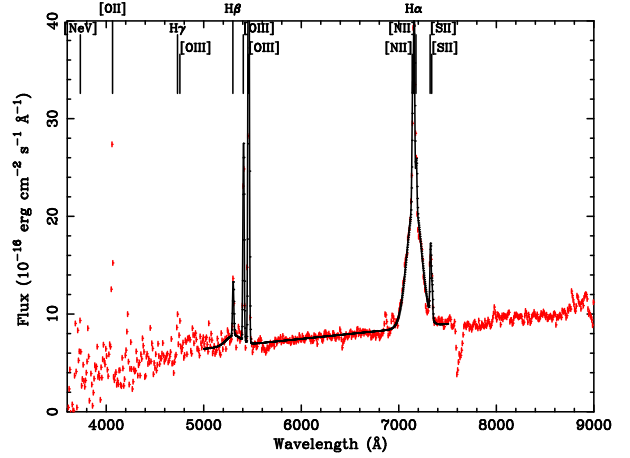}
   \caption{FOS optical spectrum for RXJ1331, along with the
   best fit model (over the 5000-7500\AA\ range) including a linear continuum
   and narrow Gaussian lines for the main emission lines, along with an
   additional broad Gaussian component for H$\alpha$ and H$\beta$ (see text for
   details).}
   \label{278-010-opt}
\end{figure}

The optical spectrum was taken from the RIXOS project. It was observed using
the Faint Object Spectrograph (FOS) on the INT in February 
1994 (see Mason et
al. 2000 for details). The spectrum is shown in Fig. \ref{278-010-opt}, along
with the best fit model to the 5000-7500\AA\ region including a linear
continuum, narrow Gaussian lines for H$\beta$, [OIII]$\lambda\lambda$4959,5007,
H$\alpha$, [NII]$\lambda\lambda$6584,6548 and [SII]$\lambda\lambda$6731,6717,
and additional broad Gaussian lines for H$\alpha$ and H$\beta$. The region used
to estimate the error bars was 5800-6400\AA. The fit is reasonable, with no
obvious residuals around H$\alpha$ or H$\beta$. The broad component of H$\beta$
does not look very compelling to the eye, but it improves the fit very
significantly ($\Delta\chi^2=196$ for 298 d.o.f. with just one more
parameter). We therefore classify this source as a Sy1.8, instead of the Sy1.9
classification given in Mason et al. (2000). In addition to the above lines,
[OII], H$\gamma$ and perhaps [NeV] can be distinguished in the spectrum.

The Balmer decrement for the narrow H lines is $3.47\pm0.16$, or
$E(B-V)=0.10\pm0.04$ (with the Galactic reddening of 0.03 already
subtracted). The NLR is only very slightly reddened. In contrast, the Balmer
decrement for the broad H lines is $12\pm0.7$, or $E(B-V)=1.10\pm0.07$~mag
(which corresponds to $N_H=(6.4\pm0.4)\times 10^{21}$~cm$^{-2}$, more than 50\%
higher than the value inferred from the X-ray spectrum, see section
\ref{278-010XSp}). The BLR appears therefore to be substantially reddened.

The [OIII]5007 flux (which will be used to check if the source is Compton thick
in section \ref{278-010XSp}) has been determined using the NLR Balmer decrement
to correct the [OIII]5007 line intensity for reddening. Following Bassani et
al. (1999) and Pappa et al. (2001), we correct the observed [OIII] flux by a
factor [(H$\alpha$/H$\beta$)$_{\rm NLR}$/3]$^{2.94}$, obtaining
$(1.6\pm0.3)\times10^{-13}$ erg cm$^{-2}$ s$^{-1}$. Because our spectrum was
taken through a narrow slit with relatively inaccurate absolute photometry, we
have applied a correction to the [OIII]5007 flux. We obtained a correction
factor of $0.53\pm0.07$ by comparing the flux in the spectrum with our CCD
photometry (R=16.45$\pm$0.06, Carrera, Page \& Stevens 2003), and hence obtain
a final value of $(0.85\pm0.19)\times10^{-13}$ erg cm$^{-2}$ s$^{-1}$ for the
[OIII]5007 line flux.

\subsection{RXJ213807.61-423614.3}
\label{031-001OSp}

\begin{figure}
   \centering
   \includegraphics[angle=-90,width=8cm]{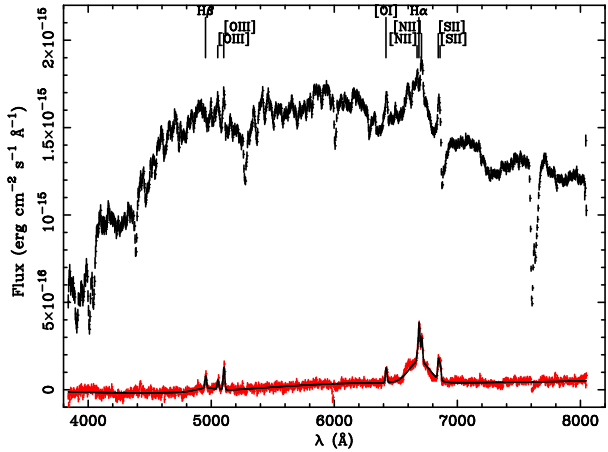}
   \caption{Optical spectra for RXJ2138: the brightest one is for the nuclear
plus galaxy emission, while the fainter one is the galaxy-subtracted nuclear
spectrum. We also show the best fit model to the latter, over the whole
wavelength range of spectrum.  This model includes a five node spline fit to
the continuum, narrow 
Gaussian lines for the main emission lines, and
additional broad 
Gaussians for H$\alpha$ and H$\beta$ (see text).}
 \label{031-001-opt}
\end{figure}

The optical spectrum is from the European Southern Observatory 3.6m
Telescope. It was obtained in photometric conditions through a 1.5$''$ slit
using EFOSC2 with the 300 lines mm$^{-1}$ grating, yielding 20\AA\ resolution
(FWHM measured from arc lines through the same slit). The continuum shape
(Fig. \ref{031-001-opt}) is clearly that of a galaxy at $z=0.019$ with NaI, MgI
and CaII absorption lines. A few emission lines can be seen over this
continuum, namely [SII]$\lambda\lambda$6732,6717 (badly affected by an
atmospheric band), [NII]6584, H$\alpha$, [OIII]$\lambda\lambda$ 4959,5007, and
probably H$\beta$, H$\gamma$ and [OIII]4363. To get a better estimate of the
emission line parameters, we have subtracted the off-nuclear galaxy spectrum to
leave only the nuclear component. Now broad H$\alpha$ can be easily seen, as
well as narrow H$\beta$. We have fitted to this latter spectrum a 5-point
spline for the continuum, narrow 
Gaussians for H$\beta$, [OIII]$\lambda\lambda$
4959,5007, [OI]6300, [NII]$\lambda\lambda$6548,6584, H$\alpha$, and
[SII]$\lambda\lambda$6732,6717, and a broad 
Gaussian for H$\alpha$. Adding a
second broad 
Gaussian at the observed position of H$\beta$ results in
$\Delta\chi^2=35$ (1004 d.o.f. in total), with an F-test probability of
$10^{-7}$, so H$\beta$ is detected with high statistical confidence. This
source is then a Sy1.8.

The Balmer decrement of the NLR is $3.0\pm0.4$, or in terms of reddening,
$E(B-V)=-0.05\pm0.12$ (Galactic value of 0.04 already subtracted). The NLR is
essentially unreddened.  In contrast, the BLR shows a Balmer decrement of
$7\pm1$, or $E(B-V)=0.60\pm0.14$.

Since the observing conditions were photometric, and the spectrum was obtained
through a slit which was 1.5 times the seeing (1 arcsec), we make no
photometric correction to the observed [OIII]5007 flux. The [OIII]5007 line
arises only from the central $\sim$arcsecond of the galaxy because its
intensity is similar in the nuclear and galaxy+nuclear spectrum 
(Fig. \ref{031-001-opt}).

\subsection{RXJ163054.25+781105.1}
\label{122-013OSp}

\begin{figure}
   \centering
   \includegraphics[angle=-90,width=8cm]{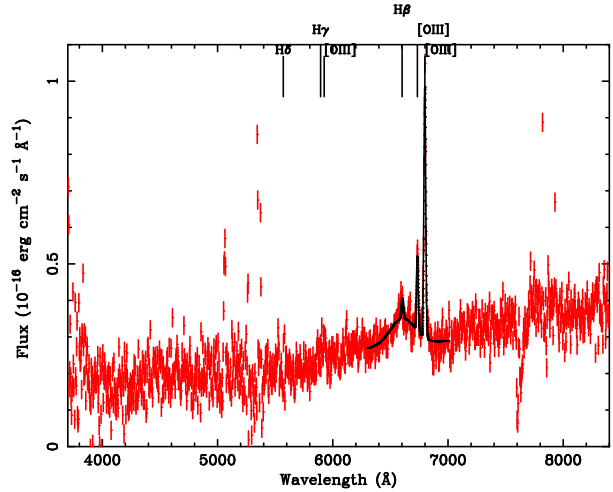}
   \caption{Optical spectrum for RXJ1630, along with the best
fit over the 6300-7000\AA\ range to a linear continuum, narrow H$\beta$/[OIII]
and broad H$\beta$.}
   \label{122-013-opt}
\end{figure}

This optical spectrum is from the RIXOS survey. It was taken in April 1993 with
the ISIS spectrograph at the WHT (see Mason et al. 2000 for details). We show
the optical spectrum for this source in Fig. \ref{122-013-opt}, along with the
best fit over the 6000-7000\AA\ range to a linear continuum with Gaussian
narrow emission lines for [OIII]$\lambda\lambda$4959,5007 and H$\beta$, and a
broad Gaussian component for H$\beta$ as well. The range used to estimate the
error bars was 6000-6500\AA. Although the centre of the Gaussian representing
narrow H$\beta$ does not coincide exactly with the peak of the spectral hump,
both components of the H$\beta$ line are significant. Hence we classify this
object as a Sy1.5.

The expected position of the H$\alpha$ line is outside the wavelength coverage
of this spectrum, and therefore we cannot measure the Balmer decrement. We can
however estimate the amount of reddening from the continuum shape, as explained
in section \ref{278-010OSp}. Matching the Francis et al. (1991) QSO continuum
to this spectrum requires either $N_H\sim(4\pm1)\times 10^{21}$~cm$^{-2}$ (much
larger than the column density inferred from the X--ray spectrum, see section
\ref{122-013XSp}) and very little galaxy contribution, or $N_H\sim(0\pm1)
\times 10^{21}$~cm$^{-2}$ and $\sim$90\% galaxy contribution. Taking into
account the presence of galactic absorption lines at the redshift of the source
in the optical spectrum, we have chosen this second solution.

To obtain an absolute flux for [OIII]5007 we apply an additional correction
factor of $1.9\pm0.4$ based on optical photometry ($R=18.73\pm0.09$), and
extrapolating the spectral continuum towards the red using a straight
line. This results in an absolute [OIII]5007 line flux of
$(0.26\pm0.05)\times10^{-14}$ erg cm$^{-2}$ s$^{-1}$.

   \begin{table*}
      \centering
      \caption[]{Log of \XMM{} observations. The column labelled Target shows
      both the target name as it appears in the \XMM{} observation log and our
      shortened names. OBS\_ID is the unique \XMM{} observation ID number. Date
      is the observation date. Texp/Tobs show the final exposure time available
      for each instrument after cleaning of bad intervals, over the total
      instrument ``on'' time.}
	 \label{XObs}
  \begin{tabular}{llrrrr}
            \hline
            \noalign{\smallskip}
         &       &    &\multicolumn{3}{c}{Texp/Tobs (s)}\\
   Target&OBS\_ID&Date&pn&MOS1&MOS2\\
            \noalign{\smallskip}
            \hline
            \noalign{\smallskip}

   278-010/RXJ1331 & 0061940101 & Jan  3, 2001 & 4174/4651 & 
6969/7039
&  6983/7043 \\
   031-001/RXJ2138 & 0061940201 & Jun  1, 2001 &1259/10485 &
3719/13010
& 3719/13017 \\
   122-013/RXJ1630 & 0061940301 & Sep 20, 2001 &   -/-     & 
4937/5003
&  4943/5003 \\
   122-013/RXJ1630 & 0061940901 & Apr 11, 2002 &  1559/7337 & 
4203/9326
&  4214/9332 \\

            \noalign{\smallskip}
            \hline
         \end{tabular}

\end{table*}

\section[]{\XMM{} observations and data}
\label{XSp}

The spectra reported here come from four different \XMM{} observations (see
Table \ref{XObs}) of three different targets. The last target (RXJ1630) was
observed twice because in the first observation the {\tt EPIC pn} instrument
(the prime instrument of the observation) was off. Most of the observing time
for RXJ1238 was lost due to high background caused by a high flux of 
soft protons.

We processed all the {\tt EPIC} data using {\tt SAS v 5.3.3}.  To assemble the
source spectra, events in circular regions (36, 20 and 25~arcsec radii for
RXJ1331, RXJ2138, and RXJ1630 respectively) around the X--ray source positions
were used. These radii were chosen such that the source counts were
significantly above the background, so as to maximise the signal to noise ratio
of the spectra. Response and effective area files were constructed using the
{\tt SAS} tasks {rmfgen} and {arfgen} respectively for each source in each
instrument and each observation. Background spectra were extracted from several
nearby bright-source-free circular regions (avoiding chip gaps in pn).  The
spectra were constructed from single and double events in pn (pattern$\le 4$),
and singles, doubles and triples in the MOS (pattern$\le 12$).

The spectra from different instruments/exposures for each source were coadded
using our own code which also combines the response
files. The background spectra were also coadded using the same
code, in a way that preserves the statistical properties of the sample, taking
into account the different source and background areas for each spectrum. 
A full description of the recipe used to combine the spectra, backgrounds and 
response matrices is given in Page, Davis and Salvi 
(2003). The
coadded spectra were binned to have at least 10 counts per bin (RXJ2138 and
RXJ1630) or 20 counts per bin (RXJ1331 which has more counts) to allow use of
the $\chi^{2}$ statistic. Only channels with nominal
energy between 0.2 and 12 keV were used in the fits.

We have fitted several models (with increasing complexity) to each source,
checking at each step the significance of the improvement over the previous fit
using the F-test. We have started with a single powerlaw spectrum with
absorption by cold gas, with column density fixed at the Galactic value,
(models {\tt phabs} and {\tt zpower} in {\tt xspec}). We have then added a
second absorption component at the redshift of the source with free column
density ({\tt zphabs}), to test for intrinsic absorption.  Finally, we have
added a Gaussian line with a rest-frame energy of 6.4~keV, and its velocity
width fixed to 0, to represent a narrow Fe K$\alpha$ emission line.

The best fit values are shown in Table \ref{XSpResults}, and each source is
discussed separately below. All uncertainties quoted are 2$\sigma$ for one
interesting parameter.

\begin{table*}
 \centering
 \caption[]{Summary of X-ray spectral fits. The values marked with
 an asterisk (*) are kept fixed in the fit. The values in brackets under the
 Source column are the intrinsic column densities from fits to the \Rosat{}
 spectra in units of $10^{20}$ cm$^{-2}$. F.P. is the F-test probability. The
 fluxes ($S_S$ and $S_H$ in the 0.5-2 and 2-10~keV bands, respectively) are
 corrected for Galactic absorption. The luminosities are corrected both for
 Galactic and intrinsic absorption.}
\label{XSpResults}

  \begin{tabular}{cclllrrrrrr}
  \hline
  \noalign{\smallskip}
  Source & $z$ & $N_{H,Gal}$ &  $N_H$ & $\Gamma$ & 
$\chi^2/\nu$ & F.P. & $S_S/S_H$ &$L_{\rm 2-10\,keV}$
&$\epsilon_0$ & EW\\
         &     & \multicolumn{2}{c}{($10^{20}$ cm$^{-2}$)} & &
& (\%) & ($10^{-14}$~cgs)&($10^{42}$~cgs) & (keV) & (eV)\\

   \noalign{\smallskip}
   \hline
   \noalign{\smallskip}

  RXJ1331 & 0.090* & 1.9* & -  &
0.70$^{+0.03}_{-0.04}$   &
543.24/92 &       &       &  \\
$[63^{+9}_{-8}]$          &        &        & 41$^{+7}_{-5}$ &
1.49$^{+0.09}_{-0.10}$ &
 96.95/91 & 100.0 &       &   \\
   &        &        & $42^{+5}_{-6}$       &
$1.51\pm0.10$ &
 88.22/90 & 99.6 & 35/176 & 35 &  6.4*       &
$180\pm120$ \\

   \noalign{\smallskip}
   \hline
   \noalign{\smallskip}

 RXJ2138  & 0.019* & 2.6* & -     & 0.8$\pm0.2$      
       &
 22.33/15 &   \\
$[60\pm30]$          &        &        & 29$^{+45}_{-11}$ &
1.3$^{+0.6}_{-0.3}$    &
 13.19/14 & 99.2 \\
   &        &        & 29$^{+48}_{-12}$ &
1.3$^{+0.6}_{-0.4}$    &
 13.19/13 & -    & 4/21 & 0.2 & 6.4*          &   0$^{+800}$
\\

   \noalign{\smallskip}
   \hline
   \noalign{\smallskip}

 RXJ1630  & 0.358* & 4.1* & -  & 1.3$\pm0.11$ 
         &
 39.97/48 \\
$[33^{+16}_{-17}]$ &        &        & 12$^{+11}_{-8}$ &
1.6$\pm0.2$     
      &
 30.85/47 & 99.95 \\
    &        &        & 13$^{+11}_{-8}$ & 1.6$\pm0.2$     
      &
 29.73/46 & 80.6   & 7/19 &  72 & 6.4*        &
160$^{+320}_{-160}$\\

  \noalign{\smallskip}
  \hline

\end{tabular}
\end{table*}

\subsection{RXJ1331}
\label{278-010XSp}

\begin{figure}
   \centering
   \includegraphics[angle=-90,width=8cm]{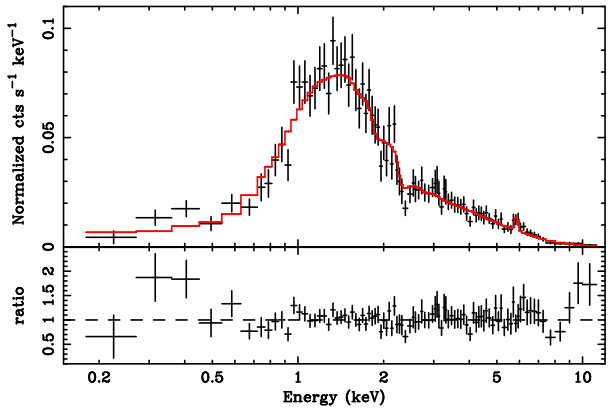}
   \caption{Spectrum for RXJ1331 (crosses), along with the best fit model
   (stepped line, including Galactic absorption, redshifted absorption,
   powerlaw continuum, and a broad emission line). The ratios between the data
   and the best fit model values are also shown.}
   \label{278-010-fit}
\end{figure}

This is the source with the highest flux in our sample, and hence the one with
the best X--ray spectrum. The single powerlaw model is a very bad fit. The
significance of introducing absorption intrinsic to the source is $\sim$100\%,
with an intrinsic column density which is at least $\sim$20 times the Galactic
value.  The fit improved (at 99.25\%) with the introduction of a narrow
Gaussian line at a rest energy of~6.4 keV.  The EW is $180\pm120$~eV, which is
consistent with values found in nearby, bright Sy1-1.5 (Nandra et al. 1997).
This fit is shown in Fig. \ref{278-010-fit}, and its parameters reported in
Table \ref{XSpResults}. We also tried fits in which the width and central
energy of the Fe line were free parameters, but these were not significantly
better fits to the data ($< 99\%$ significance according to the F-test).

Some residuals can be seen between 0.2 and 0.5~keV. We have tried both a
partial covering model and a blackbody soft excess to reduce them, but the
improvement in the fit was not significant, and obvious residuals still
remained in place.

We have tried to fit the X--ray continuum using an ionized absorber. There is
no improvement over the neutral gas absorption model (significance $\leq
70$\%). The best fit powerlaw slope is very similar to the fit with a cold
absorber, ($1.52\pm0.11$), although the column density is slightly higher
($(50^{+6}_{-11})\times 10^{20}$~cm$^{-2}$). However, the ionization parameter
is very low ($\xi\sim L/N_eR^2\sim0.012^{+0.022}_{-0.012}$). With such an
ionization parameter, the abundant elements are at most singly ionized (Kallman
\& McCray 1982) and hence the absorption detected in RXJ1331 is from cold
material.

The residuals above 7~keV prompted us to try a reflection model (Magdziarz
\& Zdziarski 1995) for this object. The fit to this model with
only the relative reflection free, and a 
Gaussian line with rest energy fixed
to 6.4~keV and width fixed to 0 was very good, with $\chi^2=84.09$ for 89
d.o.f.. In F-test probability terms this corresponds to an improvement of
99.82\% over the cold absorption with no Gaussian line. The EW of the line was
$120\pm110$~eV. However, this fit predicted that the reflection component was
9$^{+5}_{-2}$ times the direct component, i.e., this source would be reflection
dominated. To check this possibility, we calculated the position of RXJ1331 in
the Bassani et al. (1999) diagram. Using the absorption and calibration
corrected [OIII]5007 line flux value given in section \ref{278-010OSp}, and the
source hard flux from Table \ref{XSpResults}, we get $S_H/F_{[OIII]}=21\pm5$,
well into the Compton thin regime in that diagram.

Therefore, the best fit model for this source with the present data is an
intrinsically absorbed (but Compton thin) powerlaw with an Fe~K$\alpha$
emission line. The column density obtained from the Balmer decrement of the BLR
(under the assumption of standard gas-to-dust ratio), is a factor of $>$1.5
larger than the value required by the X--ray spectrum. A non-standard
gas-to-dust ratio would in principle alleviate this apparent
contradiction.

\subsection{RXJ2138}

With only 17 bins after grouping, this spectrum does not warrant very complex
models. A single powerlaw with only Galactic absorption gives a reasonable fit
with a very hard powerlaw slope ($\Gamma=0.8$). The introduction of intrinsic
cold absorption improves the fit at $>99\%$, with a column density of between
6
and $\sim$30 times the Galactic value, though still with a rather hard powerlaw
slope ($\Gamma=1.3$).

\begin{figure}
   \centering
   \includegraphics[angle=-90,width=8cm]{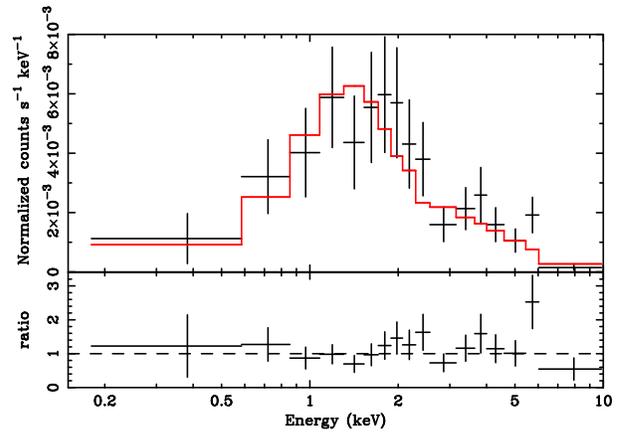}
   \caption{Spectrum for RXJ2138 (crosses), along with the best fit model
   (stepped line, including Galactic absorption, redshifted absorption, and a
   powerlaw continuum). The ratios between the data and the best fit model
   values are also shown.}
   \label{031-001-fit}
\end{figure}

The spectrum and model are shown in Fig. \ref{031-001-fit}. An attempt to
introduce an additional narrow ($\sigma=0$) Gaussian line component does not
improve the fit, and  leaves the other parameters
practically unchanged. The upper limit to the EW of the line is 800~eV. 

We have used the [OIII] line flux given in section
\ref{031-001OSp} and the source hard flux from Table \ref{XSpResults}
to get $S_H/F_{[OIII]}=102\pm8$.
This alone situates this source well into the Compton thin regime in
the Bassani et al. (1999) diagram.  This source is thus best fitted by an
intrinsically hard powerlaw with Compton thin intrinsic absorption. Further
components are not required by the data.  The absorbing column density deduced
from the BLR Balmer decrement is of the order of that required by the X--ray
data.

\subsection{RXJ1630}
\label{122-013XSp}

\begin{figure}
   \centering
   \includegraphics[angle=-90,width=8cm]{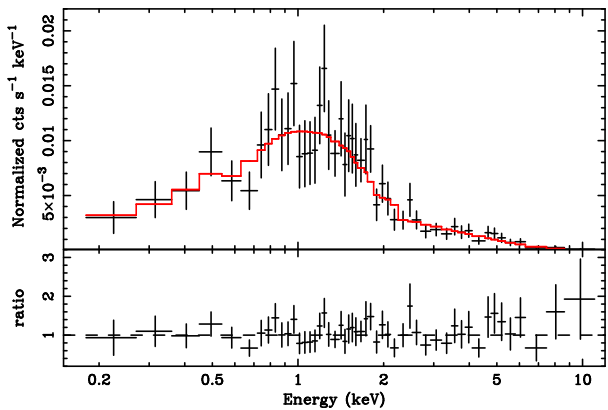}
   \caption{Spectrum for RXJ1630 (crosses), along with the best fit model
   (stepped line, including Galactic absorption, redshifted absorption, and a
   powerlaw continuum). The ratios between the data and the best fit model
   values are also shown.}
  \label{122-013-fit}
\end{figure}

A fit to a single powerlaw looks reasonable in $\chi^2$ terms, and gives rise
to a hard powerlaw. The fit improves at $>$99.9\% if 
intrinsic absorption is
included, with an inferred column density which is between 2 and 6 times the
Galactic value. This model is the one shown in Fig. \ref{122-013-fit}.

As in the case of RXJ2138, the introduction of a narrow Gaussian at 6.4 keV
does not significantly improve the fit, and the other parameters remain
practically unchanged.  The upper limit to its EW is $500$~eV. 
We have checked whether this source is Compton thick. From section
\ref{122-013OSp} and table \ref{XSpResults}, its $S_H/F_{[OIII]}=72\pm14$ using
the observed [OIII] flux.
The source is again well away
from the Compton thick region in the Bassani et al. (1999) diagram.

Again, the best model fit to this source is a powerlaw with Compton thin
intrinsic absorption. The data do not warrant the introduction of more
sophisticated models. The column density deduced from the broad-band optical
reddening is within the errors, of the order of the value fitted to the X--ray
spectrum.

\section{Discussion}
\label{discussion}

The optical spectra of our three AGN define them as intermediate-type Seyferts,
with signs of mild (if any) NLR obscuration, but strong BLR obscuration (at
least in two of them).  Their X--ray spectra are well fitted by relatively hard
powerlaws, absorbed by moderate columns ($\sim$a few $\times10^{21}$
cm$^{-2}$).  But, how do the parameters from the X--ray fits compare to those
of other AGN at similar fluxes? How does the optical obscuration compare to the
X--ray absorption?

\subsection{Slope and column density distribution}

The intrinsic absorbing column densities from the \XMM{} spectra are between
1.5 and 3 times smaller than those deduced from \Rosat{} data (see Table
\ref{XSpResults} and Page, Mittaz \& Carrera 2001). This is probably mainly
due to the assumption of $\Gamma=2$ we had to adopt in that paper because of
the limited number of independent bins in the \Rosat{} spectra. In all three
cases the \XMM{} spectra show that our sources have intrinsically somewhat
flatter spectra than $\Gamma=2$.

To examine how our sources relate to the wider AGN population, we have compared
the powerlaw slopes and column densities of our sources to those of other AGN
with broadband ($\sim 0.5-10$~keV) X--ray spectra.

Significant samples of bright Seyferts have been observed with {\sl ASCA} and
form a good benchmark with which to compare our sources.  George et al. (1998)
studied the {\sl ASCA} X--ray spectra of a sample of 18 nearby Sy1-1.5. Two
objects in their sample showed significant instrinsic absorption: IC4329A, has
$N_H\sim4\times10^{21}$~cm$^{-2}$ (of the order of the $N_H$ found for
RXJ1331), and NGC4151 has a larger $N_H$ of up to a few $\times
10^{22}$~cm$^{-2}$ (see also Schurch \& Warwick 2002). Only 3 of these 18
Sy1-1.5 have best-fit power law slopes which are as hard or harder than those
of our \Rosat{} selected sources.  Our hard sources appear to have flatter
slopes and much higher absorption than unobscured Sy1-1.5 of equal or higher
X-ray luminosities.  However, photoelectric absorption is more frequently
discernable in the X--ray spectra of Seyferts with higher levels of optical
obscuration. In a sample of 25 bright Sy1.9-2 observed with {\sl ASCA}, Turner
et al. (1997) found significant cold absorption in 14 objects, of which 12 are
more heavily absorbed that our \Rosat{} selected objects. In contrast to the
Sy1-1.5, almost half of these sources (11/25) have best-fit spectral slopes as
hard, or harder, than our sources.  Our hard sources seem to have similar
slopes but lower absorption than obscured Sy1.9-2 of similar X-ray
luminosities.

\begin{figure}
   \centering
   \includegraphics[angle=-90,width=8cm]{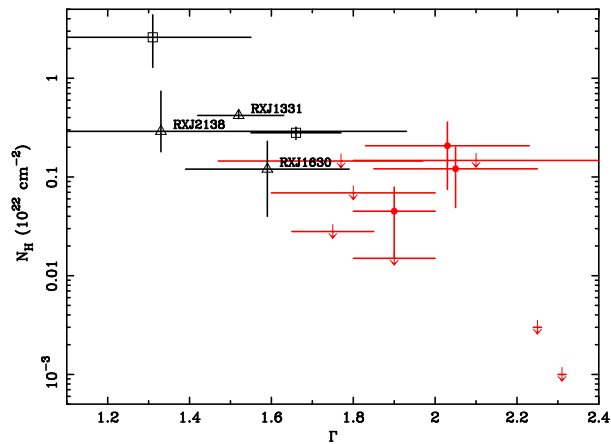}
   \caption{Intrinsic (redshfited) column density versus powerlaw slope for our
   hard BLAGN (triangles), and for the 2 BLAGN with significant absorption in
   Piconcelli et al. (2002) (squares). The AXIS BLAGN with $S_{\rm
   2-10\,keV}>10^{-13}{\rm cgs}$ are also shown, either with solid circles
(if
   the 99\% lower limit in column density is $>0$), or with arrows pointing
   down (at the top end of the 99\% confidence limits on their column
   densities).  The 90\% error bars in the last two samples have been
   transformed to 95\% assuming Gaussian statistics.  }
   \label{gnh_BLAGN}
\end{figure}

At 2-10~keV flux levels more similar to our three \Rosat{} selected sources, we
can draw a comparison sample from the serendipitous \XMM{} sources of Mateos et
al. (2003a), based on the AXIS project (Barcons et al. 2002). In that sample
there are 9 broad line AGN with 2-10~keV fluxes $\geq 10^{-13}$ cgs showing
broad optical/UV emission lines whose spectra have been modelled with a
powerlaw and intrinsic cold absorption (Mateos et al. 2003b). We show in
Fig. \ref{gnh_BLAGN} the best fit parameters for those sources, along with
those for our three \Rosat{} selected sources.  Three of the AXIS broad line
AGN show significant absorption, with similar column densities to those found
in our sources, but softer power law slopes than our three objects. On the
other hand, the two broad line AGN in the 2-10~keV selected sample of \XMM{}
sources from Piconcelli et al. (2002), for which there is evidence for
intrinsic absorption, have similar powerlaw slopes, and similar or higher
column densities to our sources (see also Fig. \ref{gnh_BLAGN}).

Inevitably, the comparison is subject to small number statistics, compounded by
the fact that none of the samples discussed (including ours) can be said to be
complete in a statistical sense. It is therefore not clear whether the
hard-spectrum \Rosat{} survey has selected sources with a different (flatter)
distribution of power law slopes to the 2-10 keV population as a whole.
However, it does appear that objects with intrinsic column densities of a few
$\times 10^{21}$~cm$^{-2}$ are an important part of the moderately bright
($\geq 10^{-13}$ cgs) 2-10 keV AGN population.

\subsection{Optical obscuration and X--ray absorption}

Our three objects show intrinsic absorption in their X--ray spectra,
well above that expected from material in our Galaxy. Furthermore, the very low
Balmer decrement in the NLR of RXJ1331 and RXJ2138 imply that this absorption
is intrinsic to the nuclei of these objects, and not due to extra-nuclear
material in their host galaxies.

The Balmer decrement on the BLR of RXJ2138 is larger than we would expect from
the best fit X--ray column density (if we assume a Galactic gas-to-dust ratio
and extinction curve), although the two measurements are compatible because of
the relatively large uncertainty on the X--ray column density. Under this
assumption, the optical reddening from the continuum shape of RXJ1630 and its
best fit X--ray column density are also compatible (within the errors).
This is not the case for RXJ1331, in which the Balmer decrement from the BLR
corresponds to a cold gas column density 1.5 times higher than that deduced
from fits to its X--ray spectrum, under that assumption.

An apparent excess of optical redenning compared to X--ray photoelectric
absorption, as observed in RXJ1331, has already been observed in several other
AGN. For example, the Sy1.8/1.9 galaxy H1320+551 (Barcons, Carrera \&
Ceballos
2003)
has a BLR Balmer decrement ((H$\alpha$/H$\beta$)$_{\rm BLR}>27$), but no
signs
of significant intrinsic absorption in its Compton thin X--ray spectrum. Pappa
et al. (2001) also find a couple of cases of optically obscured Sy2 galaxies
with no signs of X--ray absorption. These authors propose three possible
reasons for the mismatch of optical obscuration and X--ray absorption: (a) the
BLR does not exist or is intrinsically reddened; (b) the sources are Compton
thick, and the flux below 10 keV is due to scattered or host galaxy emission;
(c) a dusty warm absorber reddens the BLR but does not affect much the X--ray
properties.

We have already discarded possibility (b) for all three of our AGN, based on
the Bassani et al. (1999) diagram. We can also discard (c) for RXJ1331, as a
warm absorber is not consistent with the data.  RXJ1331 is therefore an
intriguing object. It has a small reddening of the NLR (situating the absorbing
material closer to the nucleus than the NLR), no evidence for a warm absorber,
and a strong mismatch between the optical BLR Balmer decrement and X--ray
absorption. Could it be then that some component of the large BLR Balmer
decrement is intrinsic to the BLR and is not due to optical redenning along the
line of sight?

To answer this question, we have compared the properties of RXJ1331 with those
of the sample of Seyfert galaxies of Ward et al. (1988), who find a good
correlation between the Balmer decrement and the 2-10~keV to H$\beta$ ratio
(see appendix \ref{appendix}). RXJ1331 has a Balmer decrement $\sim$0.4~dex
above what would be expected from its 2-10~keV to H$\beta$ ratio, and the Ward
et al., correlation.  Changing the gas-to-dust ratio would not bring RXJ1331
into line with the Seyfert galaxies in Ward et al. (1988): the reddening vector
in Fig. 4 of Ward et al. (1988) is parallel to their observed correlation.
This might suggest that RXJ1331 is therefore similar to, though much less
extreme than, H1320+551, which shows no significant absorption to either its
X--ray or optical continua despite its very large Balmer decrement.

However, the non-simultaneity of our X-ray observation (2001) and optical
spectrum (1994) could also be responsible for the discrepancy. RXJ1331 was
discovered serendipitously in the \Rosat{} observation with ROR number
rp701034n00 (done on July 18, 1992), with a countrate in the 0.1 to 2.4~keV
band of 0.037$\pm$0.002 and hardness ratios HR1=0.97$\pm$0.04 and
HR2=0.68$\pm$0.05. It also appears in the \Rosat{} All Sky Survey (performed in
the first half year of the ROSAT mission in 1990-1991) Bright Source Catalogue
(RASSBSC, Voges et al. 1999) with a countrate of 0.10$\pm$0.02 in the same
band, and hardness ratios HR1=1.00$\pm$0.14 and HR2=0.65$\pm$0.20.  For
comparison, we have taken the best fit model to the \XMM{} spectrum of RXJ1331
and used {\tt xspec} to calculate the expected count rates for \Rosat{},
obtaining a total countrate of 0.025$\pm$0.002, and hardness ratios HR1=0.98
and HR2=0.65. RXJ1331 thus presents remarkably stable X-ray spectral properties
over a 10 year period, but has varied in intensity such that it was a factor 4
brighter when observed in 1990/1991 than in 2001. If RXJ1331 was as bright in
X-rays in February 1994 when its optical spectrum was recorded as it was during
the \Rosat{} All Sky Survey, it is relatively consistent with the Ward et al
(1988) sample, and its Balmer decrement can be explained by a higher than
Galactic dust-to-gas ratio. If instead, it was at a similar X-ray flux level to
those observed in 1992 and 2001, it is a more complex object, similar to,
though much less extreme than, H1320+551.

We have found that the three X--ray brightest broad line AGN from the
hard-spectrum \Rosat{} survey have optical reddening similar to, or larger
than, would be expected for an absorber with a Galactic gas-to-dust ratio and
reddening law.  This finding is particularly interesting because it is contrary
to what has been found for the majority of X--ray absorbed AGN studied so
far. For example, in our own survey (Page, Mittaz
\& Carrera 2000) we found a total of 13 $z>1$ hard spectrum luminous
BLAGN of which 10 have $\log(N_H/cm^{-2})\geq22$. If ``effective'' gas-to-dust
ratios as large as the ones we have found for RXJ1331, RXJ1630 and RXJ2138 were
present in those high redshift AGN, the amount of optical obscuration would
be sufficient to completely block the BLR ($E(B-V)\sim7-13$ for their
mean $\log(N_H/{\rm cm}^{-2})\sim22.29$). This is certainly not the case,
because broad emission lines are detected in all of them.  An equally good
demonstration of this point is the study of Maiolino et al. (2001), who
constructed a sample of AGN with measurements of both optical reddening and
X--ray photoelectric absorption. The $E(B-V)/N_H$ value was found to be
significantly lower than Galactic in 16 out of the 19 objects studied. The
other 3 objects in their study were all low luminosity objects (2-10~keV
luminositiy $<10^{42}$ cgs). Our \XMM{} observations show that $\ge$Galactic
$E(B-V)/N_H$ values {\em can} be found in more luminous objects (e.g. RXJ1630
has a 2-10~keV luminosity of $7\times 10^{43}$ cgs), although this may be much
rarer in such sources.

It appears that the luminous AGN which are abundant at high redshift cannot
simply be scaled-up versions of the low redshift AGN presented here. There must
be some other ingredient, such us a luminosity dependence or evolutionary
effect in the gast-to-dust contents of the absorber which gives rise to the
different absorption characteristics in high and low redshift AGN.

\section{Conclusions}
\label{conclusions}

We have presented X--ray spectra from \XMM{} of the three brightest AGN
exhibiting broad optical emission lines (all three are intermediate Seyferts)
from the sample of hard spectrum \Rosat{} sources (Page, Mittaz \& Carrera
2000, 2001). The X--ray spectra of all three sources are well fitted by
powerlaws ($\Gamma\sim1.5$), absorbed by moderate amounts of intrinsic nuclear
cold material ($N_{\rm H}\sim$a few $\times10^{21}$ cm$^{-2}$). Similarly
absorbed sources are an important part of the $S_{\rm 2-10\,keV}\geq
10^{-13}$~cgs AGN population, although the three sources studied here appear to
have harder intrinsic power law slopes than the majority of AGN at this flux
level.

The equivalent width of the narrow emission line at about 6.4~keV found in
RXJ1331 is typical of other radio-quiet Compton-thin Seyferts.

Detailed analysis of our optical spectroscopic data confirm the classification
of these sources as intermediate-type Seyfert galaxies (Sy1.5-Sy1.9). For the
two objects in which both H$\alpha$ and H$\beta$ are visible (RXJ1331 and
RXJ2138), the NLR is shown to be almost free of reddening, while the BLR is
significantly reddened. The column density necessary to produce this effect is
about 1.5 times that inferred from the X--ray absorption in RXJ1331, and of the
order of it in RXJ2138 and RXJ1630 (in this last case from its broad band
optical spectrum), if standard Galactic gas-to-dust ratios are assumed.

None of the three sources are Compton-thick.  The X--ray data for RXJ1331
require that the absorber is cold, allowing us to rule out the presence of dust
embedded in a warm absorber in this source.  

These three low redshift broad line AGN from our sample of hard sources show a
ratio of optical extinction to X--ray absorption which is similar to, or larger
than, the interstellar medium of our own Galaxy. This cannot be the case for
the high luminosity, high redshift, broad line AGN in our sample, because the
broad lines would be completely obscured at the X--ray column densities
($>10^{22}$~cm$^{-2}$) observed in these sources. 

To explain the whole population of absorbed AGN, their effective gas-to-dust
ratio must show a large variety, perhaps depending on luminosity,
evolving with redshift, or showing geometries different from those proposed by
the unified AGN model.

\begin{acknowledgements}
We thank the anonymous referee for useful remarks. The work reported
herein is based partly on observations obtained with \XMM{}, an ESA science
mission with instruments and contributions directly funded by ESA member states
and the USA (NASA). This research was based partly on observations collected at
the European Sourhern Observatory, Chile, ESO No.  62.O-0659.  The WHT and INT
telescopes are operated on the island of La Palma by the Issac Newton Group of
Telescopes in the Spanish Observatorio del Roque de Los Muchachos of the
Instituto de Astrof\'\i {}sica de Canarias. Partial financial support for this
work was provided by the Spanish Ministry of Science and Technology under
project AYA2000-1690.
\end{acknowledgements}

\appendix

\section{Estimating the optical obscuration}
\label{appendix}

When available, we have used the Balmer decrements (H$\alpha$/H$\beta$) to
calculate the extinction. The standard unreddened Balmer decrement is $({\rm
H\alpha/H\beta})=3$ (Osterbrock 1989), as expected for case B recombination and
optically thin gas. This is expected to hold for the NLR, but there are several
theoretical reasons why case B recombination might not apply to the BLR,
because of the high density of the clouds in this region. Collisional,
self-absorption and radiative transport effects can affect the Balmer
decrement, which can have values between about 1 and 20, depending on the exact
(and largely unknown) physical state of the matter in the BLR clouds, as shown
by many theoretical works and calculations (e.g. Mushotzky \& Ferland 1984,
Netzer, Elitzur \& Ferland 1985, Rees et al. 1989, Ferguson \& Ferland
1997).

To estimate the typical intrinsic Balmer decrement in AGN we have analyzed the
region between 4600 and 7000 \AA\ of several composite optical QSO spectra,
following the same procedure we have followed for the spectra of our sources,
as outlined above. The templates have BLR Balmer decrement values around 3
(FIRST Bright QSO Survey -Brotherton et al. 2001-: $2.53\pm0.04$, SDSS -Vanden
Berk et al. 2001-: $2.38\pm0.04$, Large Bright QSO Sample and others -Francis
2003-: $3.36\pm0.09$), in any case well below the most extreme values in the
above theoretical calculations.

In a different approach, Ward et al. (1988) have measured the average intrinsic
unreddened H$\alpha$/H$\beta$ ratio in the BLR of a sample of 46 AGN (including
Sy1 to Sy1.9, broad-line radio galaxies and quasars) to be 3.5, independent of
any atomic physics assumptions. They found a good linear corretion in the log
space (see their Fig. 4) between the BLR Balmer decrement, and the ratio of
2-10~keV luminosity $L_{\rm 2-10\,keV}$ (practically unaffected by Compton thin
absorption), to H$\beta$ luminosity (strongly dependent on absorption). This
led them to suggest that the Balmer decrement is determined by nuclear
reddening, rather than being intrinsic to the BLR. By fitting a straight line
to this correlation, they find that the AGN reddening law between about 4900
and 6600\AA\ is similar to that in our Galaxy, and they relate the intrinsic
$L_{\rm 2-10\,keV}$/H$\beta$ ratio to the intrinsic Balmer decrement. They
determine the intrinsic $L_{\rm 2-10\,keV}$/H$\beta$ ratio from a subsample of
their sources deemed to be subject to very little reddening, obtaining finally
a value of H$\alpha$/H$\beta$=3.5 (see Ward et al. 1988 for details). We have
repeated their analysis including the uncertainties in the fitted parameters,
obtaining H$\alpha$/H$\beta$=3.43$\pm$0.19, again very similar to the standard
value of 3.

We will therefore use H$\alpha$/H$\beta$=3.43$\pm$0.19 as an estimate of the
intrinsic unreddened Balmer decrement value in the BLR of X-ray AGN, since this
value has been obtained from a sample of X-ray AGN with a similar X-ray
luminosity range as our sources (see Table \ref{XSpResults}, and Table 1 of
Ward et al. 1988), and independently of any atomic physics assumptions.

We have also compared the broad-band optical spectral shape of RXJ1630 (for
which no BLR Balmer decrement can be obtained) with that of the composite QSO
spectrum from Francis (2003), with some contribution from a galaxy continuum
(in our case we have tried an ES0 from the Coleman, Wu \& Weedman 1980
model). The broad-band continuum slopes of the three QSO templates cited above
are different, as is that from Francis et al. (1991), with the spectrum from
Francis (2003) being the flattest and that from Francis et al. (1991) being the
steepest. Over the rest frame range 3800 to 8500 \AA, the differences between
these two extreme examples can be parameterized as reddening by a Galactic law
with $E(B-V)=0.17$. This is the uncertainty we will assign to the reddening
determined from the broad-band optical spectral shape.

Estimating the reddening in this fashion could be misleading if the sources
happen to have an intrinsically ``red'' continuum.  Richards et al. (2003) have
studied the overall continuum and emission line properties of quasars from the
SDSS, finding that there is a population of intrinsically red quasars. They
have produced composite spectra of their ``normal'' quasars in four relative
optical magnitude bins, and estimated the slope of the optical continuum
$\alpha_\nu$ ($f(\nu)\propto\nu^{\alpha_\nu}$) to be between $\sim$-0.25 and
-0.76. Taking the line-free regions recommended by Vanden Berk et al. (2001)
(1350-1365~\AA\ and 4200-4230~\AA), we have estimated the slope of the Francis
(2003) composite spectrum to be $\alpha_\nu=-0.97$, falling with $\nu$ faster
than any of the Richards et al. (2003) ``normal'' quasars. Since the slope of
the Francis et al. (1991) spectrum is $\alpha_\nu=-0.35$, we conclude that our
estimate of the reddening from the optical continuum is conservative both in
absolute value (because we are using a spectrum with an ``intrinsically'' red
continuum), and in the assigned uncertainty (because it includes a very broad
range of continuum spectral shapes).

\end{document}